# 2006 Fragmentation of Comet 73P/Schwassmann-Wachmann 3B Observed with Subaru/Suprime-Cam

Masateru ISHIGURO

National Astronomical Observatory of Japan,

Osawa 2-21-1, Mitaka, Tokyo 181-8588, Japan

masateru.ishiguro@nao.ac.jp

Fumihiko USUI

Japan Aerospace Exploration Agency (JAXA), 3-1-1 Yoshinodai, Sagamihara,

Kanagawa 229-8510, JAPAN

Yuki SARUGAKU

Kiso Observatory, Institute of Astronomy, University of Tokyo,

Mitake 10762-30, Kiso, Nagano 397-0101, JAPAN

Munetaka UENO

Japan Aerospace Exploration Agency (JAXA), 3-1-1 Yoshinodai, Sagamihara,

Kanagawa 229-8510, JAPAN

Please send communications, proofs and offprint requests to:

Masateru Ishiguro, Dr.

National Astronomical Observatory of Japan,

Osawa 2-21-1, Mitaka, Tokyo 181-8588. Japan

E-mail: masateru.ishiguro@nao.ac.jp

# ABSTRACT

The fragmentation of the split comet 73P/Schwassmann-Wachmann 3 B was observed with the prime-focus camera Suprime-Cam attached to the Subaru 8.2-m telescope. The fragmentation revealed dozens of miniature comets (Fuse et al. 2007). We analyzed the Subaru/Suprime-Cam images, detecting no fewer than 154 mini-comets, mostly extending to the southwest. Three were close to the projected orbit of fragment B. We applied synchrone–syndyne analysis, modified for rocket effect analysis, to the mini-fragment spatial distribution. We found that most of these mini-comets were ejected from fragment B by an outburst occurring around 1 April 2006, and three fragments on the leading side of nucleus B could have been released sunward on the previous return. Several fragments might have been released by successive outbursts around 24 April and 2 May 2006. The ratio of the rocket force to solar gravity was 7 to 23 times larger than that exerted on fragment B. No significant color variation was found. The mean color index, V - R = 0.50 ± 0.07, was slightly redder than that of the Sun and similar to that of the largest fragment, C, which suggests that these mini-fragments were detected mainly through sunlight reflected by dust particles and materials on the nuclei. We examined the surface brightness profiles of all detected fragments and estimated the sizes of 154 fragments. We found that the radius of these mini-fragments was in the 5- to 108-m range (equivalent size of Tunguska impactor). The power-law index of the differential size distribution was $q = -3.34 \pm 0.05$. Based on this size distribution, we found that about 1–10% of the mass of fragment B was lost in the April 2006 outbursts. Modeling the cometary fragment dynamics (Desvoivres et al. 1999, 2000) revealed that it is likely that mini-fragments smaller than ~10–20 m could be depleted in water ice and become inactive, implying that decameter-sized comet fragments could survive against melting and remain as near-Earth objects. We attempted to detect the dust trail, which was clearly found in infrared wavelengths by Spitzer. No brightness enhancement brighter than 30.0 mag arcsec$^{-2}$ (3σ) was detected in the orbit of fragment B.

1. Introduction

73P/Schwassmann-Wachmann 3 (hereafter 73P/S-W3) is a member of the Jupiter-family comets (JFCs), orbiting the Sun with a 5.4-year period. During the apparition of 1995, 73P/S-W3 showed a huge outburst in activity. Afterward, four separate nuclei were confirmed and labeled A, B, C, and D. Of the four, fragment C was the largest and the presumed principal remnant of the original nucleus. The size of the nucleus was studied based on the standard assumption for a geometric albedo of 0.04 and a linear phase coefficient of 0.04 mag deg$^{-1}$; the upper limit of the pre-breakup radius was 1.1 km (Boehnhardt et al. 1999), and the radius of fragment C was 0.68 ± 0.04 km. Although the radius of fragment B was estimated to be 0.68 km from Hubble Space Telescope (HST) observations (Toth et al. 2003), Boehnhardt et al. (2002) established an upper limit of 0.2–0.3 km. Due to poor observing conditions, fragments A and D were not found in the 2001 apparition.

We had a good opportunity to observe these broken comet fragments during its 2006 return. From near-infrared spectroscopy, no remarkable differences between fragment B and fragment C were found (Kobayashi et al. 2007; Villanueva et al. 2006). HST photographed two fragments, B and G, on 18–20 April 2006. These images revealed several dozen mini-fragments. The Spitzer Space Telescope showed not only many fragments distributed nearly on orbit but also the debris trail between them. The debris trail (dust trail) is composed of large dust particles ejected before the last perihelion passage (Vaubaillon and Reach 2006; Reach et al. 2007). Fuse et al. (2007) made optical observations of fragment B on 3 May 2006 using the wide-field optical camera attached to the Subaru 8.2-m telescope. R-band images confirmed 58 mini-comets in the vicinity of fragment B. No fragments were found along the orbit of fragment B in their Subaru images (Fuse et al. 2007).

This spectacular Subaru image presents several concerns. We noticed that most of these fragments were distributed between the anti-solar direction from fragment B and the negative orbit velocity vector. This positioning was quite interesting because these mini-comets behaved dynamically like dust particles pressed back by solar radiation pressure against the solar gravity. No obvious dust trail was found in the Subaru optical image, even though it was clear in the Spitzer infrared image. In this study, we re-analyzed the Subaru/Suprime-Cam images using the masking method developed for the detection of faint cometary dust clouds (Ishiguro et al. 2007; Sarugaku et al. 2007; Ishiguro 2008) and constructed a comet image without contaminants (e.g., stars and galaxies). This technique enabled us to detect mini-comets brighter that 26.5 mag and diffuse light sources associated with the comet brighter than 30.0 mag arcsec$^{-2}$. Applying a unique method of examining fragment size and onset time (modified synchrones and syndynes), we examined the dynamical properties of the mini-fragments. We also studied the brightness profile of these mini-comets and deduced their sizes. Given the dynamical properties and sizes, we considered the activity of the mini-comets.

## 2. Data and Observations

### 2.1. Data

We re-analyzed the Subaru data provided by the SMOKA data server, which is operated by the Astronomy Data Center, National Astronomical Observatory of Japan (Baba et al. 2002). Observations of 73P/S-W3 were carried out by Fuse et al. (2007) using the Subaru 8.2-m telescope on Mauna Kea, Hawaii, on a single day, 3 May 2006, when fragment B was at a heliocentric distance $r_h = 1.070$ AU, a geocentric distance $\Delta = 0.112$ AU, and a solar phase angle $\alpha = 54°$. Fuse et al. (2007) used an optical CCD camera, Suprime-Cam, attached to the prime focus of Subaru. This combination provided wide-field imaging capability, 34' × 27', with a pixel resolution of 0.20" pixel$^{-1}$. The seeing was about 0.7" (FWHM), which projects to 57 km at the position of the comet. The exposure time and filters are summarized in Table 1. All comet images were taken in comet-tracking mode. Although Fuse et al. (2007) did not use the short exposure-time R-band images (10–30 s) and V-band images, we found that these were essential to (i) determining the brightness of mini-comets near fragment B, (ii) improving the signal-to-noise ratio, and (iii) identifying detected sources as having a cometary origin. Fragment B was so bright that a large area of sky near B was saturated in 120-s exposures. The signal-to-noise ratio was improved by combining all images. We used the V and R composite image to confirm the mini-fragments because the color index V-R avoids false detections (see Section 2.2). Further explanations of the Suprime-Cam and 73P/S-W3B observations appear in Miyazaki et al. (2002) and Fuse et al. (2007), respectively.

[Table 1]

### 2.2. Data Reduction

As a first step, the obtained data were reduced in the standard way with bias and flat-field corrections. These ancillary data were also provided through the SMOKA system. Because useful bias data were not obtained during the night of 3 May, we used bias frames taken on 1 May. Flux calibration was done using Landolt standard stars in the SA113 region (Landolt 1992).

The sky background was contaminated by elongated stars and galaxies because the observations were carried out in comet-tracking mode. We removed these stellar objects using the masking algorism developed for the data reduction of cometary dust trails (Sarugaku et al. 2007; Ishiguro et al. 2007; Ishiguro 2008), outlined as follows. We first made images to align the stars, because this is an effective way of detecting faint stars and galaxies. Using these images, stars were automatically detected by a source extractor program, SExtractor (Bertin and Arnouts 1996). We masked the identified objects using 18" × 6" rectangular masks. We also masked pixels identified as bad in the bias (hot pixels and lines) and flat-fielding images (pixels with sensitivity 10% higher or lower than the average). We combined the masked images with offsets to align the comet, excluding the masked pixels and shifting the background intensity to zero. Because the comet moved relative to the stars, it was possible to exclude nearly all masked pixels in the resultant composite image. Finally, we obtained V- and R-band composite images without stars. For the composite images, we used images with 120-s exposure times. Approximately 17% of the pixel values in the images were masked by this method. Therefore, the effective total exposure times were 400 s in each wavelength.

To extract the mini-comets in the composite images, we first flattened the sky background by subtracting the 23-pixel × 23-pixel (4.6" × 4.6") running median images. This is a standard image-processing technique known as "unsharp masking." The large-scale components associated with the mini-comets could be subtracted out by this method. The R-band image is shown in Fig. 1. Because we detected no significant difference in appearance between the V- and R-band subtracted images, we combined these two into single image. This VR composite image was used for the detection of faint mini-comets. We used the SExtractor again to detect the mini-comets. We found 211 mini-comet candidates in this VR composite image. The positions and magnitudes of these mini-comets were examined using the "phot" command in the IRAF/APPHOT package. We set a fixed aperture size of 2.0". This aperture gathered the light from nuclei and a portion of the light from the coma components. Assuming that fragment B was at the brightest point in the 10 s exposure image, we determined the relative positions of the mini-comets.

[Figure 1]

Of the 211 mini-comet candidates, we determined R-band magnitudes for 176 objects, V-band magnitudes for 161 objects, and both R- and V-band magnitudes for 154 objects. In Fig. 2, we compare the V- and R-band magnitudes of 154 comet candidates. A glance at Fig. 2 reveals that the V-R color indices of these 154 objects were slightly redder that of the Sun (V-R$_{Sun}$ = 0.367; Rabinowitz 1998). The mean color of the mini-comets, V-R = 0.50 ± 0.07, was similar to that of main nucleus C, V-R = 0.48 ± 0.17 (Boehnhardt et al. 1999; Lamy et al. 2004). Accordingly, we can state that at least 154 mini-comets were detected by our data analysis methods.

[Figure 2]

In Fig. 1(b), we find that some mini-fragments were elongated in the anti-solar direction. Fragment B was also elongated anti-sunward. Because these dust particles were strongly coupled to solar radiation, they could have been small particles or highly porous dust aggregates (Mukai et al. 1992; Kimura et al. 2002, 2003). We also found a disconnection near fragment B (see Fig. 1(c)). In general, this disconnection could have resulted from recent fragmentation (discussed below), accidental eruption of dust particles, or solar magnetic field reversal.

3. Results and Discussion

3.1. Observed Mini-Fragment Spatial Distribution

Figure 3 shows the position of 154 mini-comets relative to that of fragment B. As we described in Section 1, most of fragments were distributed toward the southwest (the lower right of Fig. 1). Two or three objects appeared on the trailing side of B and close to its projected orbit. Three objects appeared on the leading side of B (Fig. 1(d)). It appears that the three objects on the leading side were released with a sunward velocity component on the previous return, giving these three fragments a smaller semi-major axis than that of B, whereas most of the objects in the southwest were ejected at the current apparition and expanded by the rocket effect. Figure 4 is a histogram of the position angle of the mini-fragments, which is defined as an angular offset of the mini-comet to fragment B relative to the north celestial pole. East, south, and west correspond to position angles of 90°, 180°, and 270°, respectively. The southwest population is distributed nearly symmetrically with an average of 237.7°. A small peak appears around 217.5°, which we discuss in Section 3.3. In Fig. 5, we show the R-band magnitudes of these southwest comets with respect to the distance between B and each mini-comet. As a general trend, the bright mini-comets were located near B, whereas faint objects were distributed far from fragment B.

[Figure 3]

[Figure 4]

[Figure 5]

3.2. Interpretation: Basic Dynamics Equations

Let us consider the observed distribution of the mini-fragments from the standpoint of dynamics. The motions of "dry dust particles" and "icy comets" should differ from one another. The motion of icy comets, composed of refractory (silicates and CHON particles) and volatile (mainly $H_2O$) particles, is governed primarily by solar gravity and perturbed by the planets' gravities. When these mini-comets were in the inner solar system, the rocket effect, which induces a recoil force from the gas outflow momentum, could continuously perturb their orbits. Although the recoil force is generally referred to as the "non-gravitational force," we use the term "rocket force" in this paper to distinguish it from other non-gravitational forces, such as radiation pressure.

The equation of motion of the mini-fragments in the rectangular coordinate system can be written as

$$\mathbf{F_c} = -\frac{GM_\odot m_c}{r_h^2}\mathbf{e_r} + \sum_{i=1}^{N_p}\frac{GM_i m_c}{r_i^2}\mathbf{e_{Pi}} + m_c\left(F_1\mathbf{e_r} + F_2\mathbf{e_T} + F_3\mathbf{e_N}\right) \quad . \tag{1}$$

Here $G$ and $M_\odot$ are the gravitational constant and the mass of the Sun, respectively; $r_h$ is the distance between the Sun and comet; and $m_c$ is the mass of the mini-fragment. The second term on the right side of Eq. 1 denotes the planetary perturbations; $M_i$ is the mass of the $i$-th planet, and $r_i$ is the distance between the $i$-th planet and the comet. We considered 10 objects ($N_p = 10$: eight planets, Pluto, and the Moon). We used DE406 provided by NASA/JPL for the ephemerides of these 10 objects. $\mathbf{e_{Pi}}$ is the comet–planet unit vector. $F_1$, $F_2$, and $F_3$ represent the acceleration by the rocket effect: $F_1$ is the acceleration along the radial vector defined outward along the Sun–comet line; $F_2$, perpendicular to the radial vector in the orbit plane and toward the comet's direction of motion; and $F_3$, perpendicular to the orbit plane. $\mathbf{e_r}$, $\mathbf{e_T}$, and $\mathbf{e_N}$ are the three unit vectors along the directions of the three rocket forces that satisfy the condition $\mathbf{e_N} = \mathbf{e_r} \times \mathbf{e_T}$. The acceleration components from the rocket effect can be considered a function of heliocentric distance, conventionally written

$$F_j\left(r_h\right) = A_j\, g\left(r_h\right) \quad , \tag{2}$$

where $g(r_h)$ expresses the water ice vaporization rate as a function of heliocentric distance $r_h$

$$g\left(r_h\right) = \alpha_{TII}\left(\frac{r_h}{r_{TII}}\right)^{-m_{TII}}\left[1 + \left(\frac{r_h}{r_{TII}}\right)^{n_{TII}}\right]^{-k_{TII}} , \tag{3}$$

where $r_{TII} = 2.808$ AU, $m_{TII} = 2.15$, $n_{TII} = 5.093$, and $k_{TII} = 4.6142$. The value of $\alpha_{TII}$ is chosen such that $g(r_h = 1) = 1$, which gives $\alpha_{TII} = 0.111262$. $A_j$ in Eq. 2 is referred to as the "Type-II non-gravitational parameter" and widely applied to describe the non-gravitational motion of comets due to rocket effect (Marsden et al. 1973). The transverse component parameter $A_2$ is almost always well determined for short periodic comets because it is sensitive to the secular change in the semi-major axis, which is established using long observation intervals. $A_1$ is sensitive to perturbations of the longitude of perihelion and is often not as well determined. $A_3$ is usually the least well-determined of the three because it is sensitive to perturbations in the orbital inclination and longitude of the ascending node and neither of these perturbations is secular (Yeomans et al. 2005).

In contrast, the orbit of dust particles is determined by the solar radiation pressure, Poynting-Robertson drag, and so forth, as well as gravitational forces. The dust particles may not include ice components because of their short lifetime. Mukai (1986) studied the lifetime of water ice and found the lifetime for 1-mm particles to be less than a day at 1 AU. Ignoring the rocket force, we can express dust particle motion as

$$\mathbf{F_d} = -\frac{GM_\odot m_d}{r_h^2}\left[(1-\beta)\mathbf{e_r} - \beta\left(\frac{\dot{r}}{c}\mathbf{e_r} + \frac{\mathbf{v}}{c}\right)\right] + \sum_{i=1}^{N_p}\frac{GM_i m_d}{r_i^2}\mathbf{e_{Pi}} \quad , \tag{4}$$

where $\beta$ is the ratio of the solar radiation pressure with respect to the solar gravity and $\mathbf{v}$ is the orbital velocity of the dust particle. The first term in the bracket comes from the solar gravity reduced by the radiation pressure, whereas the second term is derived from the Poynting-Robertson drag (Burns et al. 1979). In addition to these forces from Eqs. 1 and Eq. 4, the solar wind drag and the Yarkovsky effect may perturb the orbits of dust particles and cometary fragments. As studied by Mukai and Yamamoto (1982), solar wind drag is not efficient for particles larger than 1 μm. The Yarkovsky effect is also ineffective for the short-term evolution under discussion here (less than several years). Thus, we ignored the solar wind drag and the Yarkovsky effect.

For a spherical particle of radius $a$ (cm) and mass density $\rho$ (kg m$^{-3}$), $\beta$ is defined as

$$\beta = \frac{KQ_{pr}}{\rho a} \quad , \tag{5}$$

where $K$ is the constant

$$K = \frac{3L_\odot}{16\pi G M_\odot},$$  (6)

$L_\odot$ and $c$ are the solar luminosity and speed of light, respectively. $Q_{pr}$, which is defined as $Q_{pr} = Q_{ext} - <\cos\theta> Q_{sca}$, is the radiation pressure coefficient averaged over the solar spectrum. Here, $Q_{ext}$ and $Q_{sca}$ are the efficiency factors for extinction and scattering, respectively, and $<\cos\theta>$ is the asymmetry parameter for light scattering. $Q_{pr}$ is the radiation pressure coefficient averaged over the solar spectrum (Burns et al. 1979). Assuming that the particles are compact in shape and large compared to the optical wavelength (>>0.5μm), we can fix $Q_{pr} = 1$. From observations, it is known that Jupiter-family comets eject dust particles with $\beta = 6 \times 10^{-6} - 0.2$ (Fulle 2004). It is inferred that comet brightness may be dominated by light scattered by the largest particles (i.e., smallest $\beta$; Fulle 2004, Ishiguro et al. 2007).

### 3.3. Interpreting Spatial Distribution using Modified Synchrones and Syndynes

We assumed that dust and gas emission occurred on the sunlit hemisphere of each fragment, symmetric with respect to the comet–Sun axis. This assumption is reasonable because large portions of these mini-fragments were covered with fresh icy surface when they were produced. We then expected that the rocket effect was exerted in the anti-solar direction alone. In fact, $A_1$ of fragment B is one order of magnitude larger than $A_2$ and $A_3$ (Sekanina 2005), which supports this assumption. Here, we define the ratio of rocket force acceleration with respect to the solar gravitational acceleration as

$$\begin{cases} \beta_{rkt}(r_h) = F_1(r_h)\left[\dfrac{GM_\odot}{r_h^2}\right]^{-1} \equiv \beta_{rkt,0}\, g(r_h) r_h^2 \\ F_2 = F_3 = 0 \end{cases},$$  (7)

where $\beta_{rkt,0}$ is the ratio at $r_h = 1$ AU. The position of the mini-comet parameterized by $\beta_{rkt,0}$ can be obtained by solving the following equation:

$$\mathbf{F_c} = -\frac{GM_\odot m_c}{r_h^2}\Big[1 - \beta_{rkt}(r_h)\Big]\mathbf{e_r} + \sum_{i=1}^{N_p}\frac{GM_i m_c}{r_i^2}\mathbf{e_{Pi}} \quad . \tag{8}$$

At small heliocentric distances ($r_h$ < 1.5 AU), we can assume that $\beta_{rkt}$ is independent of the heliocentric distance because $F_1$ (which is proportional to the water sublimation rate; see Eqs. 2 and 3) is approximately proportional to the solar irradiation (i.e., $\propto r_h^2$) at $r_h$ < 1.5 AU. The mathematical form of the rocket force $F_1$ at $r_h$ < 1.5 AU is similar to that of the solar radiation pressure term of the dust particle in Eq. 4. Accordingly, we can apply the synchrones and syndynes analysis for a small heliocentric distance. Figure 6 compares the synchrones and syndynes of dust particles (obtained by Eq. 4) with those of mini-comets (obtained by Eq. 5). A significant difference in the synchrones appeared when the fragment was ejected at $r_h$ > 2 AU for the above reason.

[Figure 6]

The notion of synchrones and syndynes was originally introduced to fit the observed dust tail (see e.g., Finson and Probstein 1968). In this paper we use the term "modified synchrones and syndynes" to refer to the locus of mini-fragments. In Fig. 7, we compare the positions of mini-comets and the modified synchrones and syndynes. The positions of the mini-fragments are illustrated by cross signs, and the modified synchrones and syndynes by dashed and solid lines, respectively. Most of fragments were concentrated in the range of $\beta_{rkt} = 3 \times 10^{-4}$ and $\beta_{rkt} = 1 \times 10^{-3}$. This value is about 7–23 times larger than that exerted on fragment B.

[Figure 7]

An advantage of using the modified synchrones and syndynes is that we are able to determine the onset time of fragmentation using a "snapshot." In Fig. 4, we find a prominent concentration at the position angle 237.5°. This position angle coincides with the synchrone of 26 March 2006. What happened on that date? Figure 8 shows the light curve of fragment B (obtained from Seiichi Yoshida's web site, http://www.aerith.net/). The plotted magnitude $H_\Delta$ was normalized to the geocentric distance $\Delta = 1$ AU. Two arrows, labeled O/B and SU, denote the time of expected onset on 26 March and the time of the Subaru/Suprime-Cam observation, respectively. In Fig. 8, a brightness enhancement appeared a few days after 26 March. Therefore, we can argue that most of mini-comets in the southwest were released by an outburst occurring in late March or early April. The onset time of 26 March is earlier than the generally described outburst time of 1 April (Sekanina 2007). This small discrepancy may imply that these mini-fragments were active and progressive during the early stage of the ejection and became inactive over a one-month period. Similar evidence was found for the broken comet C/1999 S4 (LINEAR): Weaver et al. (2001) estimated the separation time from the dynamical properties of 100-m mini-fragments and found that their results indicated an earlier time than the commonly accepted disruption time of C/LINEAR. In addition to the outburst on 1 April, Sekanina (2007) argued that successive outbursts occurred on 24 April and 2 May, although, in Fig. 8, we cannot find evidence for an outburst on 24 April. It is likely that the detached feature in Fig. 1(c) was the fragment produced around 2 May, and five fragments at a position angle of around 217.5° (see also the 24 April synchrone in Fig. 7(a)) might have been ejected by the outburst on 24 April (the synchrone of 24 April coincides with a position angle of 219°). It should be emphasized, however, that the outstanding single peak at a position angle of 235.5° in Fig. 4 indicates that most of the fragments in the southwest were released by the 1 April outburst.

[Figure 8]

We applied a model for the dynamics of cometary fragments to our data, following Desvoivres et al. (1999, 2000). They considered the energy balance on the surface of the icy body, given by

$$\frac{S_0}{r_h^2}\left(1 - p_v\right)\cos z = \varepsilon\sigma T^4 + L_w(T)\frac{dZ}{dt} \quad , \tag{9}$$

where $S_0$ is the solar flux at 1 AU, $p_v$ is the geometric albedo in V-band, $z$ is the zenith distance of the Sun, $\varepsilon$ is the emissivity, and $\sigma$ is the Stefan–Boltzmann constant. $T$ denotes the equilibrium temperature. Note that heat transferred to the deeper layers is neglected in this model (Desvoivres et al. 1999, 2000). The latent heat of sublimation of the water, $L_w$, is given by (Delsemme & Miller 1971)

$$L_w(T) = 2.886 \times 10^6 - 1116\left(\frac{T}{1K}\right) \quad \text{J kg}^{-1} \quad . \tag{10}$$

The sublimation rate of the water ice is given by

$$\frac{dZ}{dt} = \frac{1}{1+1/\kappa} \gamma(T) P_w(T) \sqrt{\frac{m_w}{2\pi kT}} \quad \text{kg s}^{-1} , \tag{11}$$

where $\kappa$ is the water ice-to-dust mass ratio, defined as $\kappa = \rho_w \, \rho_d^{-1}$ ($\rho_w$ and $\rho_d$ are the masses of water ice and dust particles per unit volume, respectively). $m_w$ is the molecular mass of the water. $k$ is the Boltzmann constant. $\gamma$ denotes the dimensionless sticking coefficient (Haynes et al. 1992; Enzian et al. 1997) given by

$$\gamma(T) = -2.1 \times 10^{-3}\left(\frac{T}{1K}\right) + 1.042 \quad (T > 20K) \quad . \tag{12}$$

In Eq. 11, the saturated vapor pressure of water $P_w(T)$ is given by the Clausius-Clapeyron equation:

$$P_w(T) = 3.56 \times 10^{12} \exp\left[-6141.667\left(\frac{T}{1K}\right)^{-1}\right] \quad \text{Pa} \quad . \tag{13}$$

Assuming a spherical body, the anti-solar acceleration due to sublimation of ice is given by an integral over the sunlit hemisphere:

$$F_1 = \frac{\pi f R_c^2}{m_c} \int_o^{\pi/2} \frac{dZ}{dt} v_{th} \phi \sin 2z \, dz \quad , \tag{14}$$

where $R_c$ is the radius of mini-comet, $\phi$ is the geometric correction factor, and $v_{th}$ is the mean velocity of sublimating water given by

$$v_{th} = \sqrt{\frac{8kT}{\pi m_w}} \qquad . \tag{15}$$

We introduced the fractional active area $f$ into Desvoivres's original model. $f = 1$ indicates no dust accumulation on the entire surface area. We rewrote the first equation of Eq. 7 as follows:

$$\beta_{rkt} = \frac{3r_h^2 f \int_o^{\pi/2} \frac{dZ}{dt} v_{th} \sin 2z \, dz}{4GM_{\odot}R_c \rho_c} \qquad , \tag{16}$$

where $\rho_c$ is the mass density of the mini-comet, satisfying the condition $\rho_c = \rho_w + \rho_d$. From Eq. 16, it is clear that $\beta_{rkt}$ is inversely proportional to $R_c$ and $\rho_c$, and proportional to $f$. Equation 16 suggests that smaller fragments were accelerated to higher velocities relative to fragment B, whereas larger fragments were accelerated to lower velocities, qualitatively supporting the result that the brightness of the mini-fragments decreased as the distance from fragment B increased (Fig. 5).

As discussed in Desvoivres et al. (1999), the non-gravitational acceleration by the rocket effect is not sensitive to $\kappa$ and $p_v$. We thus used $\kappa = 1$ and $p_v = 0.04$. We assumed $\varepsilon = 0.9$ and $\phi = 2/3$, following Desvoivres et al. (1999). $\beta_{rkt}$ of the fragment B around 1995 (when the initial outburst occurred) was estimated to be $4.4 \times 10^{-5}$ (Sekanina 1996), and the radius of fragment B is 680 m (Toth et al. 2003) or $< \sim 300$ m (Boehnhardt et al. 2002). By substituting $R_c = 680$ m or $\sim 300$ m and $\beta_{rkt} = 4.4 \times 10^{-5}$ into Eq. 16 we obtain

$$\begin{cases} \left[ \dfrac{\rho_c}{1000 \, \mathrm{kg\,m^{-3}}} \right]^{-1} f = 2.5 & R_c = 680m \\[4mm] \left[ \dfrac{\rho_c}{1000 \, \mathrm{kg\,m^{-3}}} \right]^{-1} f = 1.1 & R_c = 300m \end{cases} \tag{17}$$

for fragment B during the 1995 apparition. Given a mass density $\rho_c = 200–800$ kg m$^{-3}$ (Sosa and Fernandez 2008), we found $f > 0.5$ (when $R_c = 680$ m) or $0.2 < f < 0.9$ (when $R_c$ = 300 m), indicating that a large fraction of fragment B's surface was very active shortly after birth. It is reasonable to think that a large portion of the icy surface on fragment B was exposed in the 1995 breakup. Applying Eq. (16), we can calculate a typical mini-comet radius ($\beta_{rkt} = 3 \times 10^{-4}$ to $1 \times 10^{-3}$) of about 10–100 m, assuming that $f$ and $\rho_c$ of the mini-comets were the same as those of fragment B. This size estimate is roughly consistent with the results obtained by the photometry in Section 3.4. Further discussion on the size and the activity will occur in the next section.

### 3.4. Photometric Results

It is possible to derive the size of mini-comets from photometry. The magnitudes we obtained from aperture photometry in Section 2.2 must overestimate the brightness of the nuclei because of the effects of the near-nucleus coma. The rocket effect suggests the existence of sublimating ices. To examine the non-stellar nature of mini-comets, we constructed normalized radial surface brightness profiles for the mini-comets and compared them to stellar profiles obtained with sidereal tracking. The stellar profile was constructed using the exposure that was taken to bring the camera into focus. We found that the stellar profile was similar to the one-dimensional surface brightness profile of the field stars in the comet exposures, suggesting little variation in seeing over the observation period. Figure 9 shows example brightness profiles for the mini-comets and the reference star. In the graph, we show two extreme cases: "mini-comet 1," apparently the most active; and "mini-comet 3," inactive. The brightness profiles of the other mini-comets are intermediate between these two ("mini-comet 2" is the example). To deduce the size of the mini-comets, we adapted the method of Lamy et al. (1998). Assuming that the coma surface brightness decreases as $\rho^{-\eta}$, where $\rho$ is the projected distance between the line of sight and the nucleus, we estimated the flux from the nucleus. The corresponding total surface brightness distribution, $B(\rho)$, is thus given by

$$B(\rho) = \left[ k_{coma}\rho^{-\eta} + k_{nucleus}\delta(\rho) \right] \otimes PSF, \tag{18}$$

where $\delta(\rho)$ is the Dirac $\delta$ function and $\otimes$ is the convolution operator. $PSF$ is the point spread function obtained by the reference stars as mentioned above. $k_{coma}$ and $k_{nucleus}$ are the brightness scaling factors of the coma and the nucleus, respectively. Because the coma brightness of the small fragment is too faint to determine from its surface brightness, we used the typical value of the large fragment, i.e., $\eta = -2$. Using this procedure, we found that the comet's nuclei contribute >25% of the total scattering cross-section measured with a 2" aperture.

[Figure 9]

The apparent R-band magnitude of the airless body in the solar system is written following Fernandez et al. (2000) and Jewitt (2006) as

$$p_R R_C^2 = C r_h^2 \Delta^2 10^{-\frac{2}{5}\{m_R - b\alpha - m_\odot\}} \quad , \tag{19}$$

where $C$ is the constant value $2.25 \times 10^{22}$ m$^2$, $m_R$ is the R-band magnitude of the fragment's nucleus at the heliocentric distance $r_h$ and the geocentric distance $\Delta$, and $m_\odot$ is the solar R-magnitude (−27.1). The R-band albedo $p_R$ is in the range $p_R \sim p_V$ = 0.02–0.08 (Lamy et al. 2004; Tancredi 2006). Because the data were taken at the phase angle $\alpha = 54°$, the observed magnitude was highly influenced by the scattering phase function. We correct the phase effect by the linear law. The slope $b$ of the phase function is in the range 0.02–0.06 mag deg$^{-1}$ (Lamy et al. 2004; Fernandez et al. 2000). Given that $p_R = 0.04$ and $b = 0.04$ mag deg$^{-1}$, as is typical of the nuclei of short-period comets (Fernandez et al. 2005), we obtained the size of the fragment in the range of $R_c$ = 5–108 m. This is consistent with the size estimate by $\beta_{rkt}$ in Section 3.2. Uncertainty in the phase function caused a 60% error in $R_c$, whereas a 50% error in the geometric albedo translated into a 25% error in $R_c$. The uncertainty in the correction for coma contamination is at most 20%. From these uncertainties, we consider that the radius of the bare nucleus is uncertain by a factor of ~2 or less.

3.5. Total Mass Loss and Brightness Increase

Given that all fragments have the same albedo value and phase function slope, we are able to deduce the size distribution. The cumulative size distribution is shown in Fig. 10. Filled circles denote the size determined by standard assumptions for the geometric albedo and the linear phase coefficient, i.e., $p_R = 0.04$ and $b = 0.04$ mag deg$^{-1}$.

[Figure 10]

Because large fragments have a stochastic problem and small fragments have large measurement uncertainties, we fit the slope to the cumulative fragment size distribution between 12 m and 37 m. We found that the differential size distribution, defined as

$$N(R_c)dR_c = N_{ref}\left(\frac{R_c}{R_{ref}}\right)^q dR_c \quad ,$$
(20)

was $q = -3.34 \pm 0.05$. The number of fragments with a reference size $R_{ref} = 1$ m is also obtained from the above fit, and found to be $N_{ref} = (3.40 \pm 0.20) \times 10^4$. The power-law index of size distribution turns out to be similar to the index of the theoretical distribution, $q = -3.5$, expect for self-similar collision cascades as predicted by Dohnanyi (1969), but steeper than the slope of short-period comets ($q = 2.4$–2.6; Meech et al. 2004, Weissman and Lowry 2003). It is interesting that the slope is quite similar to that of the dust particles in the dust trails of 2P/Encke, 4P/Faye, 22P/Kopff, 67P/Churyumov-Gerasimenko (Ishiguro et al. 2007; Sarugaku et al. 2007; Ishiguro 2008), and dust tails (Fulle 2004; Fulle 1992; Fulle 1990). This similarity suggests that the cometary debris size distribution might be a simple power law distribution over a wide size range. Because $q > -4$, the total mass of the fragments $M_{total}$ strongly depends on the largest fragment. This enables us to deduce the total mass accurately. The total mass $M_{total}$ is given by

$$M_{total} = \frac{4}{3}\rho\pi\left[\sum_{i=1}^{N(>R_{ref})} R_i^3 + \int_{R_{min}}^{R_{ref}} N_{ref} R_{ref}^3 \left(\frac{R}{R_{ref}}\right)^{3+q} dR\right] \quad .$$
(21)

Using the parameter above, the total mass of the fragments was $1.62 \times 10^{10}$ ($\rho_c/10^3$ kg m$^{-3}$) kg for a standard geometric albedo and linear phase coefficient. Note that there should be a factor of 8 or less uncertainty due to the unknown albedo and the linear phase coefficient. Applying the $R_c = 300$–680 m of fragment B [equivalent to a mass of $0.11$–$1.32 \times 10^{12}$ ($\rho_c$ ($10^3$ kg m$^{-3}$)$^{-1}$ ) kg], we found that 1–10% of the mass was lost in the outbursts of April 2006.

We next considered the brightness change on 1 April. The magnitude of fragment B increased by $\Delta H_A \sim 3$ from $H_A \sim 14.5$ to $H_A \sim 11.5$ around 1 April (see Fig. 8). The total cross-section of the material ejected by the outburst depends on the smallest particles because $q < -3$. The cross section is given by

$$
\begin{aligned}
C_{total} &= \pi \left[ \sum_{i=1}^{N(>R_{ref})} R_i^2 + \int_{R_{min}}^{R_{ref}} N_{ref} \left( \frac{R}{R_{ref}} \right)^{2+q} dR \right] . \\
&\approx -\frac{N_{ref}}{3+q} \left( \frac{R_{min}}{R_{ref}} \right)^{3+q} \qquad (q < -3)
\end{aligned}
\tag{22}
$$

Particles with $R_{min} \sim < \lambda * 2\pi^{-1}$ will scatter optical light inefficiently because of Rayleigh scattering. Therefore, we set the minimum size $R_{min} = 1.0 \times 10^{-7}$ m (0.1 μm). If we assume that the scattering properties (i.e., albedo and phase function) of a sub-micron particle are the same as those of the decameter-sized fragments, we can deduce the brightness increase by fragmentation using Eqs. 19 and 22. We found an estimated magnitude increase of $1.3 \pm 0.6$ mag. Admittedly, the calculated flux increase is less than half the observed flux increase; nonetheless, we argue that our estimate for the power-law index is appropriate for explaining the observations because (i) the other light sources we considered in Eq. (21), such as gas emission from volatiles and scattered sunlight by secondary dust particles released from the fragments, must have been present; and (ii) the scattering properties of sub-micron dust particles contain uncertainties. It might be better to state that Eq. (21) gives a lower limit of $q$ for submicron particles; that is, if $q < -3.7$, the magnitude increase exceeds the observed increase of $\Delta H_A \sim 3$ mag. Therefore, it seems reasonable to think that the power-law index of size distribution of the comet debris is about 3.3–3.4 over a wide size range.

### 3.6. Active Area Fraction

As described above, the positions of the mini-comets relative to fragment B result from acceleration by rocket force. Because the observed separation between B and the mini-comets is the distance projected on the celestial plane, it is not easy to determine the real distance. We drew lines perpendicular to the synchrone of 1 April and derived $\beta_{\text{rkt}}$ of the fragment from the distance between B and the foot of a perpendicular. By applying Eq. 16, we obtained $f \rho_c^{-1}$. The result is shown in Fig. 11.

[Figure 11]

The left axis denotes $f \rho_c^{-1}$, and the mass density is unknown. Here, assuming that the mass density of the mini-comets is the same as 9P/Tempel 1 and 81P/Wild 2 (i.e., $\rho_c \sim$ 450 kg m$^{-3}$), we found that all fragments except one ($R_c = 100$ m) have a fractional active area $f < 1$. Perhaps fragments with $f = 1.4$ and $R_c = 100$ m have low densities ($\rho_c < 320$ kg m$^{-3}$) to fit the $f < 1$ condition. It is likely that $f \rho_c^{-1}$ decreases with decreasing radius, becoming almost constant below $\approx$10–20 m. This suggests that water ice in the surface layer within <10 m might be exhausted by sublimation. In fact, the point-source-like comet in Fig. 9 had the smallest value of $f \rho_c^{-1}$, suggesting that this comet was depleted in water ice. Because small fragments can become inactive even around 1 AU, we would suggest that such decameter-sized comet fragments could remain as near-Earth objects. Figure 11 illustrates our presumption that a fraction of near-Earth asteroids could be produced by fragmentations of comets as seen in 73P/S-W.

### 3.7. Dust Trail

The appearance of our optical image is quite different from that of the infrared-wavelength image taken by Spitzer. The infrared images showed not only the mini-comets but also the dust trails connecting each fragment (Vaubaillon and Reach 2006; Reach et al. 2007). Figure 12 shows a cut profile of the sky background perpendicular to the comet orbit. The cut profile is the result averaged more than ~20' along the projected orbit. No signal from the dust trail could be found. We put an upper limit on the surface brightness of 30.0 mag arcsec$^{-2}$ (3$\sigma$). The upper limit of the surface brightness is converted into the optical depth multiplied by albedo A($\theta$) × $\tau$, and we found that A($\theta$) × $\tau$ = 2.2 × 10$^{-12}$. A comparative study between optical and infrared observations of the dust trail would provide information about the optical properties of the dust trail.

[Figure 12]

4. Summary and Remarks

In this paper, we have described the data analysis and interpretation of the optical image of 73P/Schwassmann-Wachmann 3B taken by the Subaru/Suprime-Cam. We detected at least 154 mini-fragments whose colors were similar to those of the Sun. Except for a few fragments, they were systematically distributed toward the southwest. We applied a modified synchrone-syndyne analysis to the spatial distribution of a number of mini-fragments from the standpoint of dynamical evolution by rocket force. We found that most of these mini-comets were ejected from fragment B on 1 April 2006. This result is consistent with the evidence that the magnitude of B surged around 1 April. Three fragments on the leading side of B could have been released with a sunward velocity component on the previous return. Several fragments at a position angle of around 217.5° might have been released by an outburst around 24 April, and a detached feature near B could be related to an outburst around 2 May 2006. The ratio of rocket force to solar gravity is approximately 7–23 times larger than that exerted on B, which implies that the radii of the mini-comets are about 10–100 m. No significant color variation was found; the mean color index V-R = 0.50 ± 0.07 was slightly redder than the solar color, suggesting that these mini-comets were observable mainly through sunlight scattered by red dust particles and nuclei surfaces. We examined the brightness profiles of all detected mini-comets and deduced the sizes of 154 fragments. The power-law index of the differential size distribution was $q$ = −3.34 ± 0.05. Based on the size distribution, we found that about 1–10% of fragment B's mass was lost in outbursts occurring in 2006. No dust trail was detected along fragment B's orbit. We placed an upper limit on the surface brightness of 30.0 mag arcsec$^{-2}$ (3$\sigma$) and on the optical depth multiplied by albedo, $A(\theta) \times \tau = 2.2 \times 10^{-12}$.

We applied Desvoivres's model to derive $f \rho_c^{-1}$, including the energy balance on the icy surface and ignoring dust mantle accumulation. Although small dust particles can be lifted by escaping gas pressure, large dust particles cannot escape and gradually accumulate on the surface. This thin dust layer may significantly diminish the vaporization of water (Prialnik and Bar-Nun 1988; Rosenberg and Prialnik 2007; Prialnik et al. 2008). In fact, the surface temperature on the large portion of 9P/Tempel 1 is in good agreement with the temperature for the standard thermal model, which suggests that cometary surfaces could be covered with a dust mantle (A'Hearn et al. 2005). It should be noted that, for the model, we included an unrealistic assumption in Eqs. 9–16 that the mass density ($\rho_c$, $\rho_w$, and $\rho_d$) and the water ice-to-dust mass ratio ($\kappa$) were constant. It is natural to think that $\rho_w$ and $\kappa$ near the surface decrease with water ice sublimation. The dust density near the surface could also decrease because the sublimating water vapor pushes the dust particles into interplanetary space. The fraction of active surface area $f$ could decrease due to ice depletion. However, we could not incorporate the depletion of ice near the surface in the model (Eqs. 9–16). Studying the vaporization of water from the subsurface layer is part of our future work.

Acknowledgement

This Suprime-Cam image was taken using Subaru Telescope, and obtained from the SMOKA, which is operated by the Astronomy Data Center, National Astronomical Observatory of Japan. The visible lightcurve data were obtained from Seiichi Yoshida's web site. We sincerely thank people who are involved in these projects. We also thank Dr. T. Fuse (Subaru observatory) and his colleagues, for their effort in the data acquisition by Subaru. Also we acknowledge two anonymous referees, Prof. J. Watanabe, and Dr. T. Kasuga for their valuable comments. The ephemeris data, DE406, was provided by NASA/JPL.

p.490.

Table 1. Subaru/Supreme-Cam data we used in this paper

| UT (2006 May 03) | Filter | Exposure Time (sec.) | Number of exposure |
|---|---|---|---|
| 14:13 | R | 30 | 1 |
| 14:17-14:28 | R | 120 | 4 |
| 14:32 | R | 10 | 1 |
| 14:41-14:53 | V | 120 | 4 |

Figure Caption

Fig. 1. R-band composite image of 73P/Schwassmann-Wachmann 3B. These images are oriented in the standard fashion; that is, north is up and east is to the left. The nucleus of fragment B is masked (white area) due to the saturation. In (a), two arrows labeled N and W denote north and west; the two arrows labeled $v$ and $\odot$ denote the directions of the comet's orbital motion and anti-solar direction. The brightness scale limits and contrast were modulated to emphasize the faint fragments. (b) and (d) are enlarged views of (a) enclosed by the rectangles. (c) is obtained from a short exposure (10 s) image. All images were processed by subtracting the 23-pixel × 23-pixel (4.6" × 4.6") running median images. A disconnection (see Section 2.2) is indicated by arrows.

Fig. 2. Comparison between V-band magnitude and R-band magnitude. The average color index V-R = 0.55 is slightly redder than that of the Sun (V-R = 0.367).

Fig. 3. Observed position of the mini-fragments relative to the position of fragment B. The dashed line denotes fragment B's projected orbit.

Fig. 4. Histogram of mini-fragment position angle. The mean position angle is 237.5°. The dashed line marks the Gaussian fit. The positions of fragment B's orbit are indicated by arrows (L denotes the leading direction and T the trailing direction).

Fig. 5. R-band magnitude ($M_R$) vs. distance between the mini-fragments and B. The solid line is the fitted line, $M_R$ = 1.97 log(distance) + 14.53.

Fig. 6. Comparison of the locus of synchrones (a) and syndynes (b) for dust particles (solid lines) and mini-comets (dashed lines). (a) The synchrones are characterized by ejection times on 16 September ($r_h$ = 3.0 AU) and 22 November ($r_h$ = 2.5 AU) 2005, and 19 January ($r_h$ = 2.0 AU), 14 March ($r_h$ = 1.5 AU), 26 March ($r_h$ = 1.39 AU), 15 April ($r_h$ = 1.2 AU), and 24 April ($r_h$ = 1.14 AU) 2006, in a clockwise direction. (b) The syndynes are characterized by the parameter $\beta$ = $1 \times 10^{-4}$, $1 \times 10^{-3}$, $1 \times 10^{-2}$, $1 \times 10^{-1}$, in a clockwise direction.

Fig. 7. Comparison of the locus of the mini-fragments with modified synchrones and syndynes. (a) Thin dashed lines are modified synchrones characterized by ejection times on 16 September ($r_h$ = 3.0 AU) and 22 November ($r_h$ = 2.5 AU) 2005 and 19 January ($r_h$ = 2.0 AU), 14 March ($r_h$ = 1.5 AU), 26 March (best-fit synchrone curve of the observed position of the mini-fragments), and 15 April ($r_h$ = 1.2 AU) 2006, rotating clockwise from the closest curve to the dash-dotted line, which denotes the projected orbit of B. (b) The solid lines are modified syndynes characterized by $\beta$ = 1 × $10^{-4}$, 1 × $10^{-3}$, 1 × $10^{-2}$, 1 × $10^{-1}$, rotating clockwise from the closest curve to the dash-dotted line (projected orbit of B).

Fig. 8. Optical light curve of fragment B during the 2006 apparition. Plus signs denote the magnitude normalized to the observer distance $\Delta$ = 1 AU. O/B and SU are the time of the outburst expected through synchrone analysis (26 March) and the time of the Subaru/Supreme-Cam observations (3 May). 1, 2, and 3 denote the times of the three outbursts described by Sekanina (2007). The solid line was obtained by fitting the data between −150 days and −65 days before the perihelion passage, $H_\Delta$ = 12.7 + 15.0 log10($r_h$). The original light curve data were obtained from S. Yoshida (http://www.aerith.net/).

Fig. 9. Radial surface brightness profile of three fragments and a reference star from R-band data. Mini-comet 1 is apparently the most active, whereas mini-comet 3 is inactive. The other fragment shows an intermediate brightness profile. Three thin solid lines with slopes of −1, −2, and −3, as marked, have been included for reference.

Fig. 10. The cumulative size distribution of the mini-comets. Filled circles denote the radius determined using standard assumptions for the geometric albedo and the linear phase coefficient, i.e., $p_R$ = 0.04 and $b$ = 0.04 mag $deg^{-1}$. The dashed line shows the power-law index of the differential size distribution $q$ = −3.34, which is the result of fitting between 12 m and 37 m.

Fig. 11. $f\,\rho_c^{-1}$ vs. radius $R_c$. The solid line is the result of a running median smoothing within a window of seven nearby samples. The right axis marks the fractional active area when $\rho_c$ = 450 kg $m^{-3}$.

Fig. 12. Cut profiles perpendicular to the comet orbit. North is to the left and south is to the right.

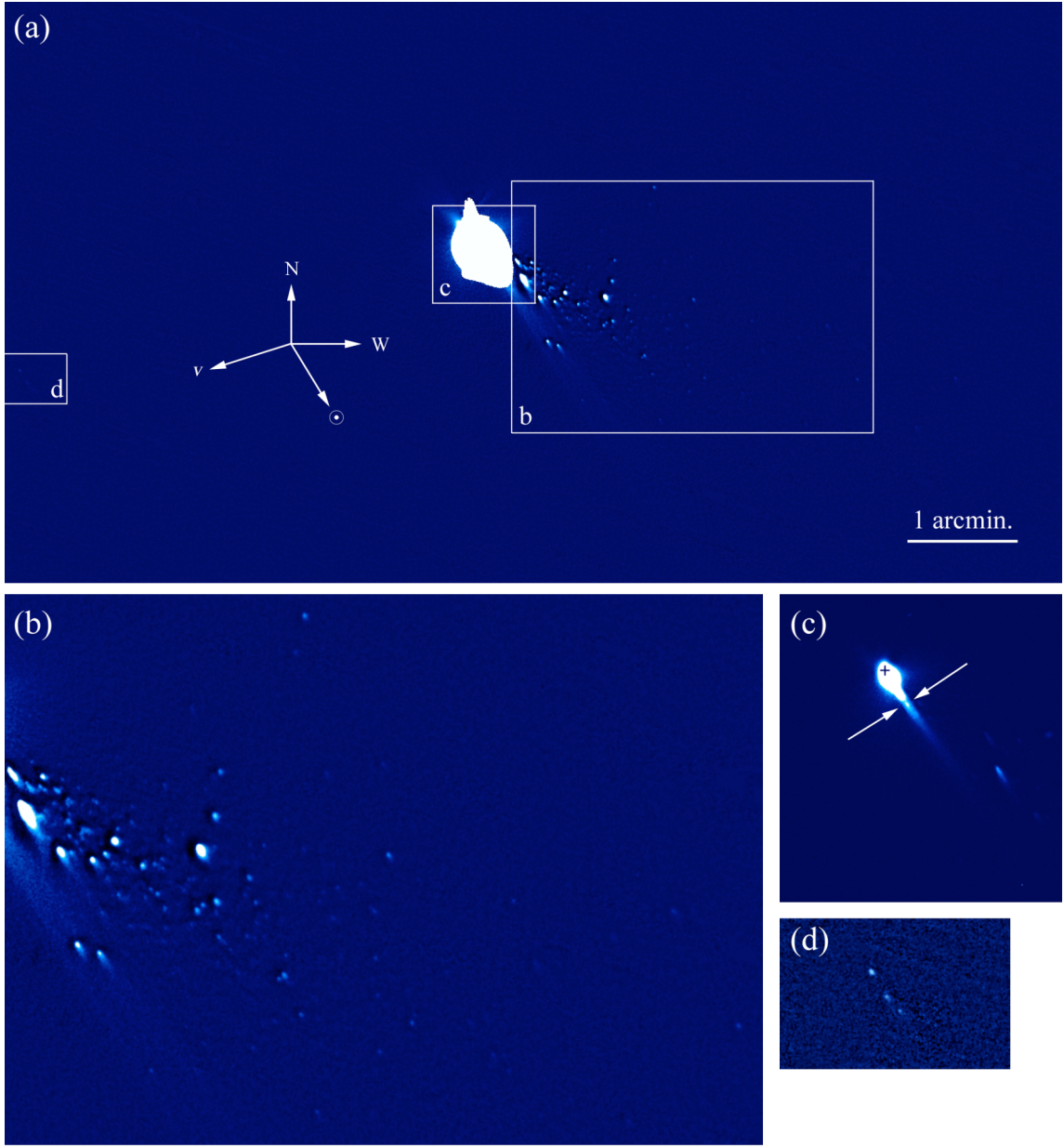

Figure 1 (Ishiguro et al.)

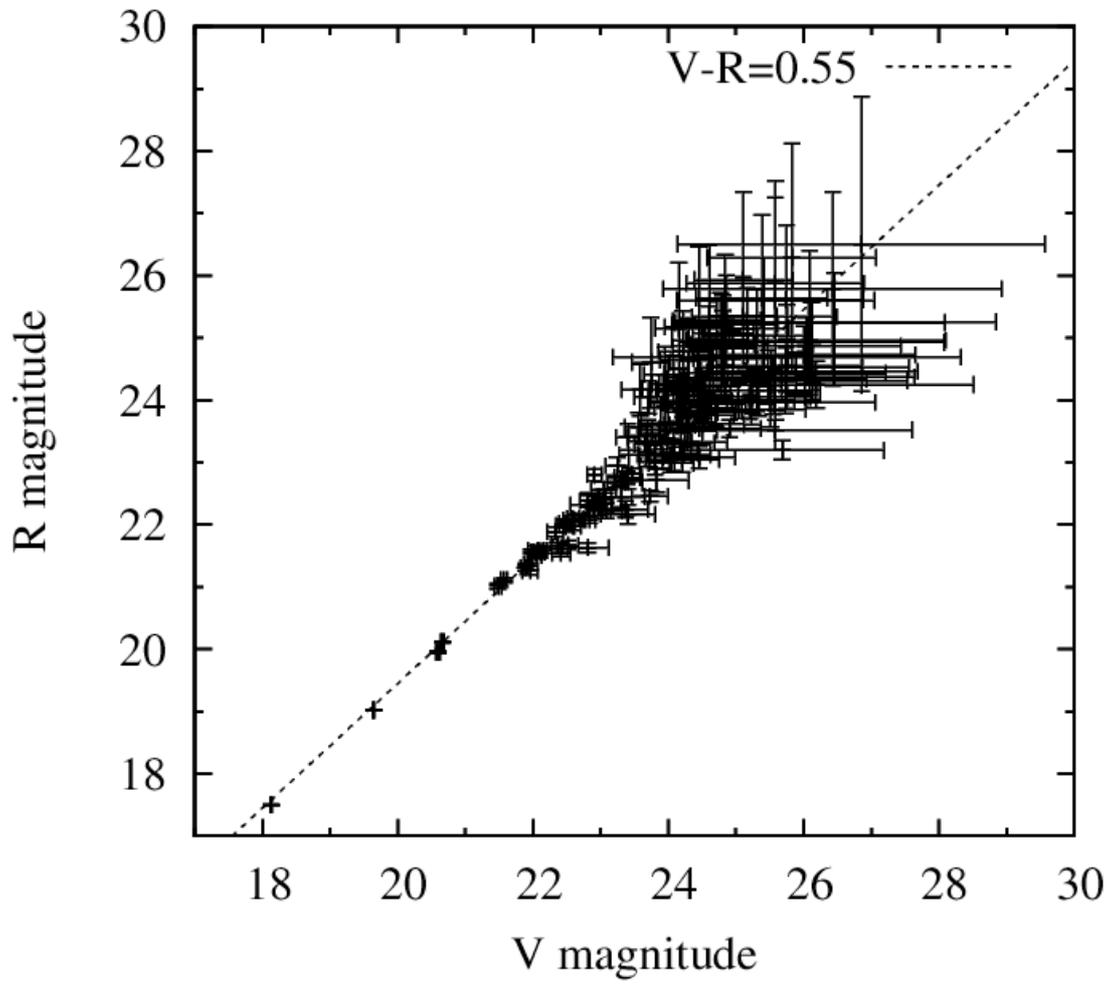

Figure 2 (Ishiguro et al.)

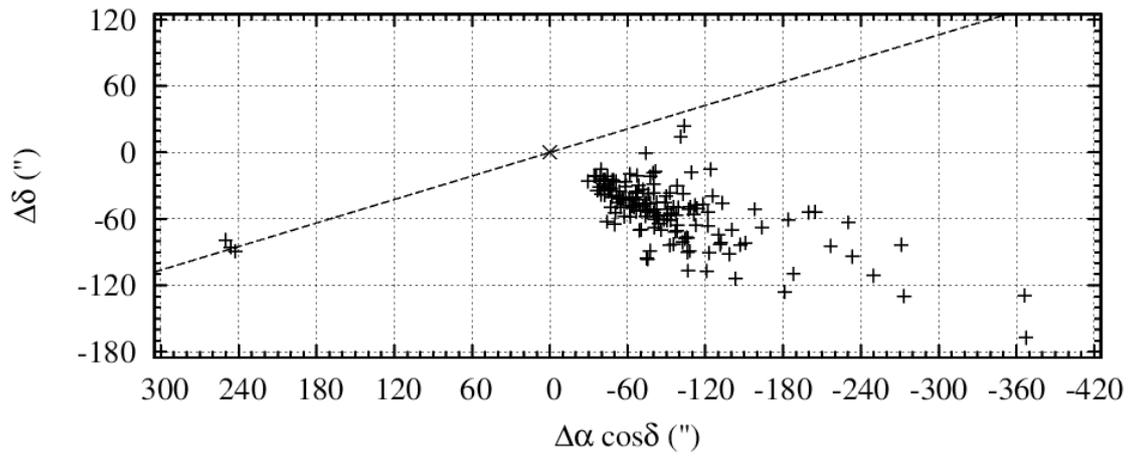

Figure 3 (Ishiguro et al.)

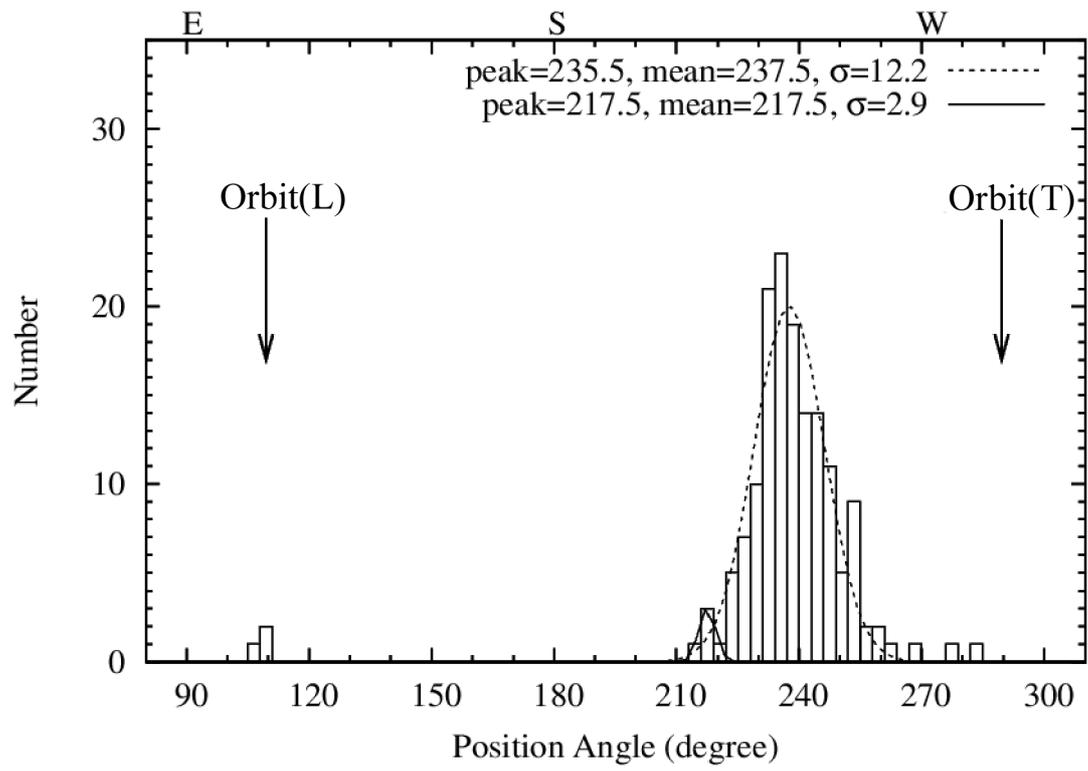

Figure 4 (Ishiguro et al.)

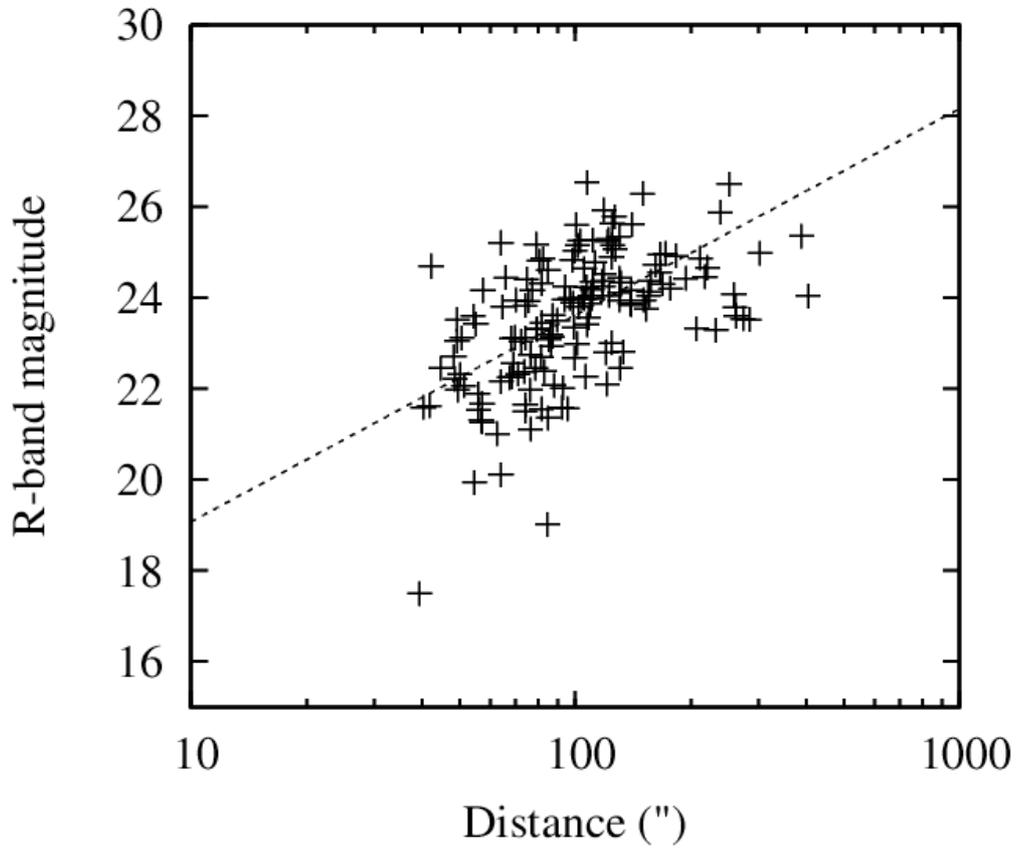

Figure 5 (Ishiguro et al.)

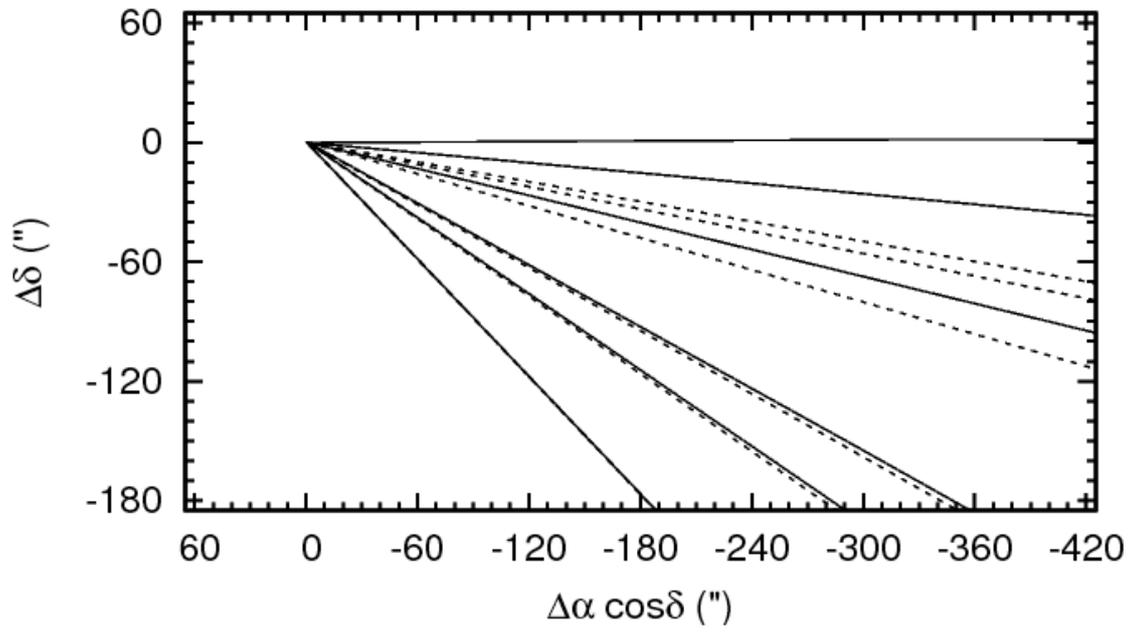

Figure 6a (Ishiguro et al.)

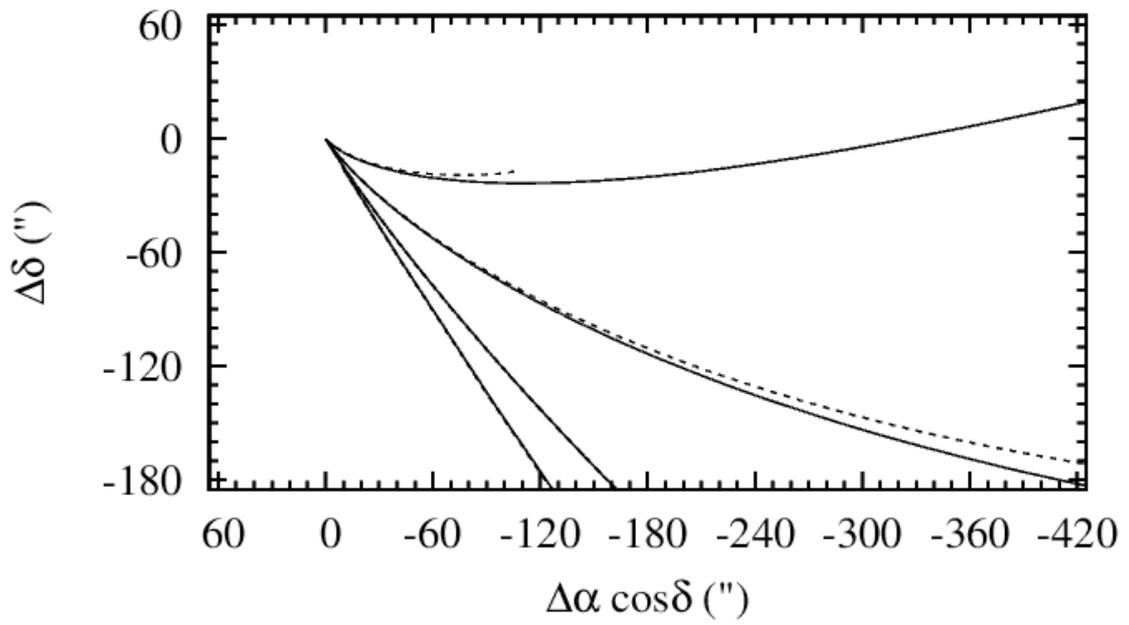

Figure 6b (Ishiguro et al.)

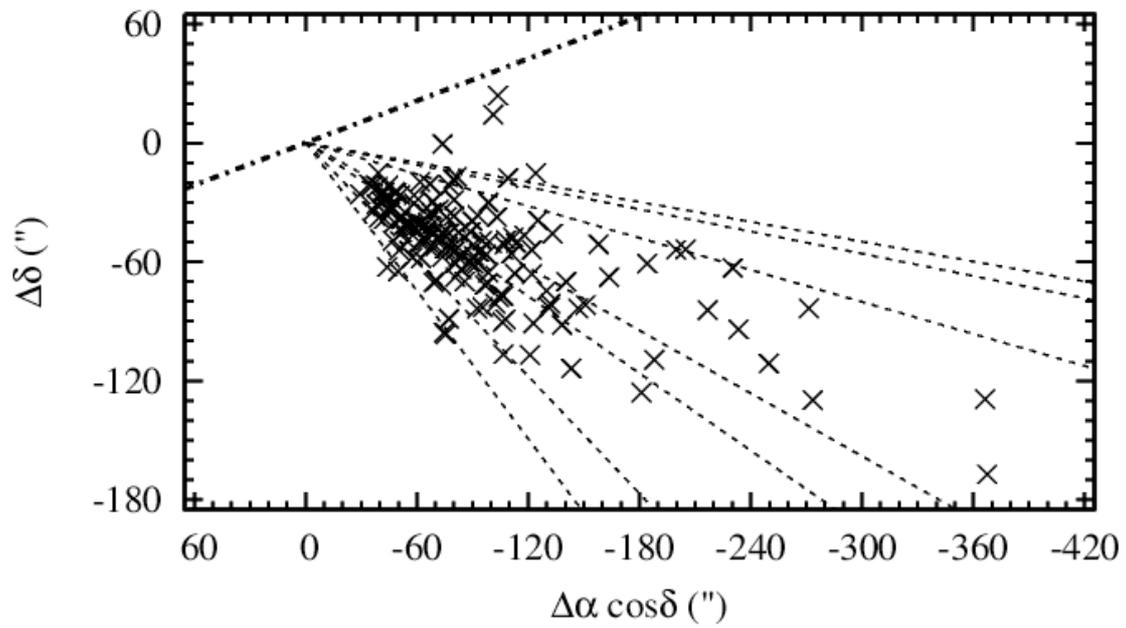

Figure 7a (Ishiguro et al.)

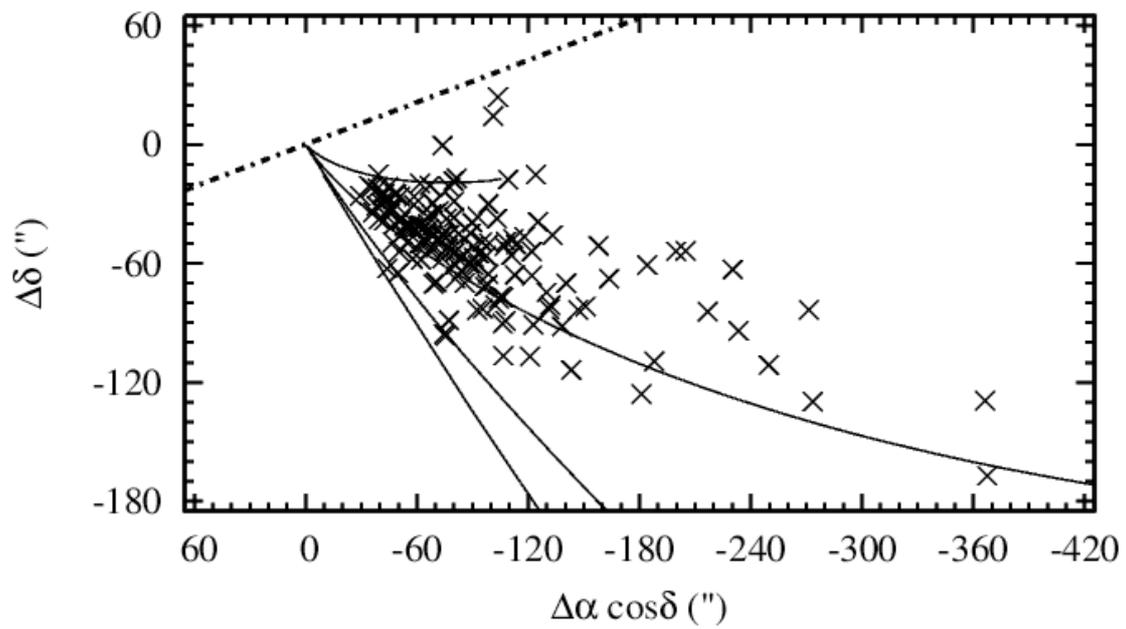

Figure 7b (Ishiguro et al.)

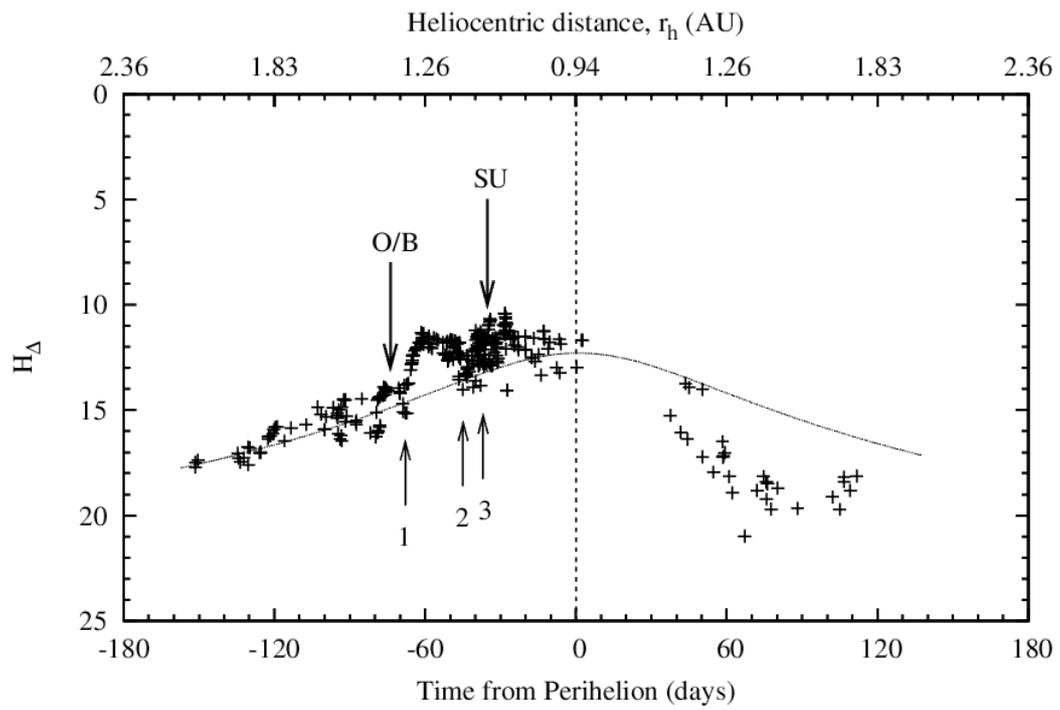

Figure 8 (Ishiguro et al.)

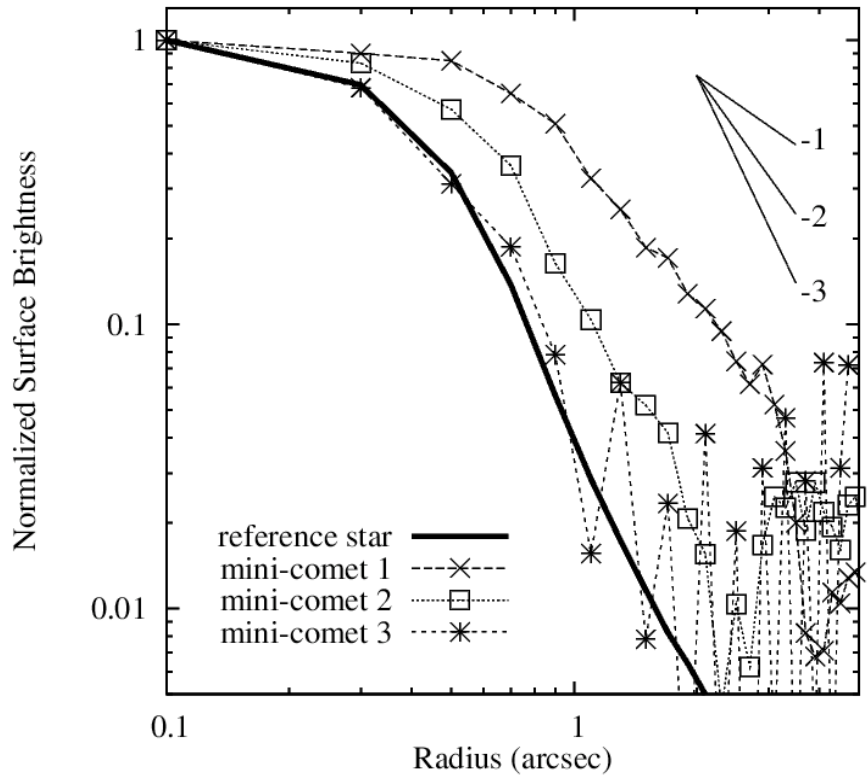

Figure 9 (Ishiguro et al.)

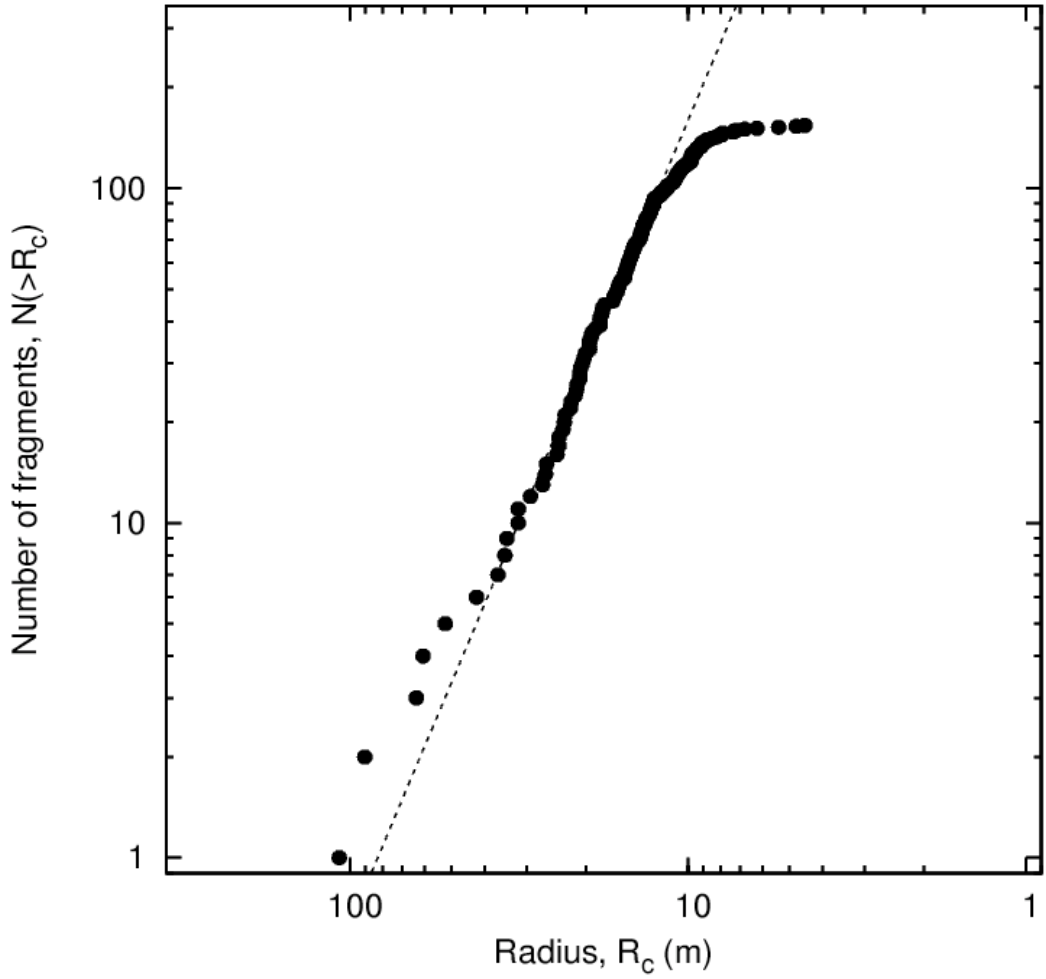

Figure 10 (Ishiguro et al.)

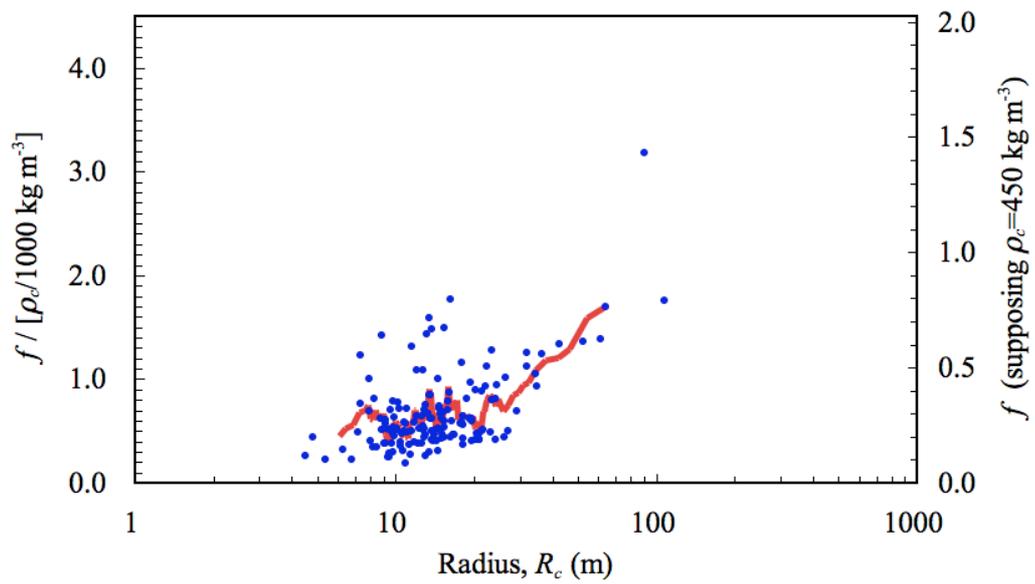

Figure 11(Ishiguro et al.)

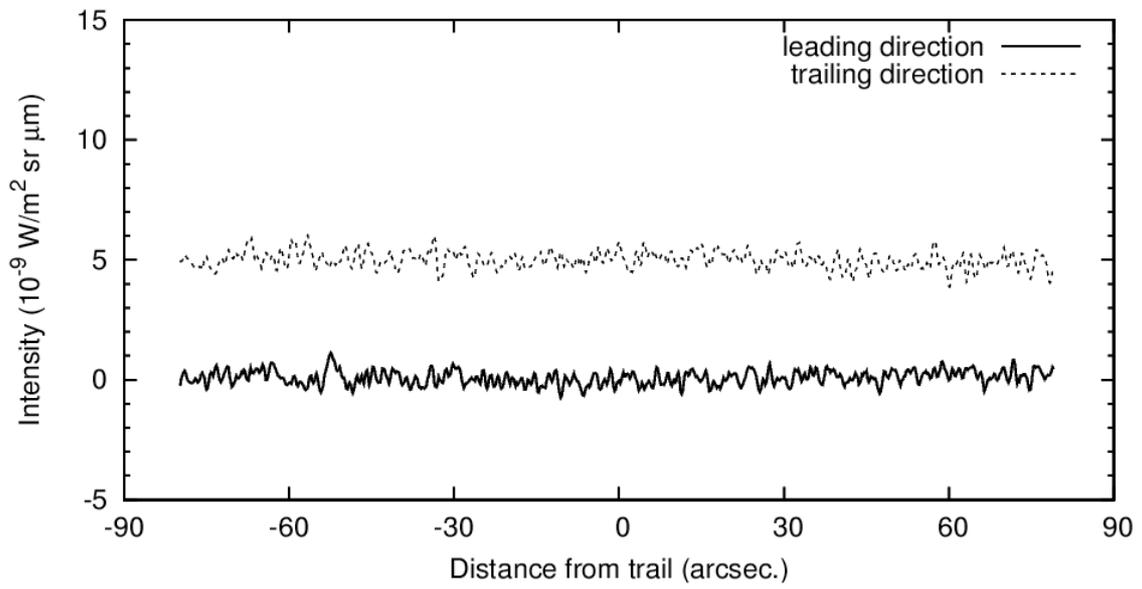

Figure 12 (Ishiguro et al.)